\documentstyle[eqsecnum,aps,epsf]{revtex} 

\begin{document}

\title{Electron exchange model potential: Application to 
  positronium-helium scattering}

\author{P. K. Biswas and Sadhan K. Adhikari}
\address{Instituto de F\'{\i}sica Te\'orica, Universidade Estadual Paulista,
01.405-900 S\~ao Paulo, S\~ao Paulo, Brazil\\}

\date{\today}
\maketitle
\begin{abstract}
The formulation of a suitable  nonlocal model potential for electron
exchange is presented, checked with electron-hydrogen and electron-helium
scattering, and applied to the study of elastic and inelastic scattering and
ionization of ortho positronium (Ps) by helium. The elastic scattering and
the $n = 2$ excitations of  Ps are investigated using a three-Ps-state
close-coupling approximation.  The higher ($n\ge 3$) excitations and
ionization of Ps atom are  treated in the framework  of Born approximation
with present exchange.
Calculations are reported of  phase shifts, and
elastic, Ps-excitation, and total cross sections.  The present  target
elastic total cross section agrees well with experimental results at thermal
to  medium energies.

{\bf PACS Number(s): 34.10.+x,  36.10.Dr} 
\end{abstract} 

\section{Introduction}
The neutral Ps beam provides a great deal of advantage over charged
projectiles 
as a probe to study the structure of atoms, molecules and surface.
Recently, there have been a great deal of interest in  positronium- (Ps-)
atom scattering due to the improvement of Ps sources and Ps beams.
Total scattering cross section of ortho Ps, which has a larger lifetime
than para Ps, have been measured for various targets 
 \cite{1,2} with an objective of understanding the Ps-interaction
dynamics with matter. Among all Ps-atom
systems, the positronium-hydrogen (Ps-H)  system is the simplest and is of
special theoretical  interest \cite{3}.  However, due to experimental
difficulties in obtaining  nascent-hydrogen atomic target there has been no
experimental study of Ps-H scattering. The next most complicated Ps-atom
system is the positronium-helium (Ps-He) system  in which  there are good
experiments on total cross section \cite{1,2}.  However, there are no
theoretical studies \cite{4a,4b,4c,4d} which can account for the measured total
cross sections of ortho Ps-He scattering . We address the present study
towards an understanding of  the measured total cross sections of Ps-He
scattering at low and medium energies using a suitably
developed  model exchange
potential.
  
The interaction of  neutral Ps atom with neutral atom or molecule is very
much
different from that of charged electron and positron with neutral targets 
\cite{11}.
In any Ps-atom scattering, the elastic and even-parity state transition
direct amplitudes to close-coupling approximation (CCA) are zero \cite{3}
due to internal charge and mass symmetry of Ps. 
 In addition, the adiabatic polarization potential is also zero and the
electron-exchange mechanism appears as the main driving force at low
energies apart from the correction expected from polarization/Van der
Waals force \cite{3}.  This was not the case for electron-impact
scattering where both the direct and exchange interactions play their
roles in determining the solution of the scattering equations. The Ps-atom
system allows the possibility for studying the effect of exchange in an
environment characteristically different from that of the electron-atom
system
due to the composite nature of Ps.  Recently, in addition to total cross section at medium to high
energies \cite{1}, thermalization of Ps in gaseous He has also been
measured
\cite{2}.  However, it is of serious concern that dynamical calculations
with the reliable and widely used static-exchange model with usual
antisymmetrization \cite{4a,4b,4c} fail severely yielding very large total
cross sections compared to the measured data \cite{1,2}, specially at low
energies.  The experiments of Refs. \cite{1,2} are consistent among
themselves. They collectively suggest a lowering trend of cross sections
from a peak at 20 eV towards lower energies. This trend is missing from
all previous published 
calculations. Moreover, due to the large error bar on the
measured cross section at 10 eV of Ref. \cite{1} and absence of data near
Ps excitation and ionization thresholds, it is not clear whether the cross
section has a minimum or not in this energy region. The present study also
addresses this feature from a theoretical point of view.

 The proper inclusion of exchange effect is a major technical obstacle in
performing dynamical calculations in complex systems \cite{S}.  The effect
of electron exchange is usually accounted for in a quantum dynamical
calculation through the antisymmerization of the wave function which
introduces nonorthogonal functions to these calculational schemes
including the usual static-exchange model.  These antisymmetrization
schemes with nonorthogonality defect lead to overcompleteness in the
Hilbert space and associated theoretical and numerical difficulties in the
CCA and related formalisms.  Moreover, when short-range (exchange) 
correlations are important, the CCA converges very slowly \cite{6a}.
Several discussions and prescriptions to remedy this defect have appeared
in the literature in connection with electron impact scattering \cite{S}. 
This problem has been overcome to some extent in electron-impact
scattering using different methodologies $-$ with essentially exact
(variational) treatment of exchange in simpler cases, with effective
correlation and suitable model potentials \cite{T} for larger targets. 
Gross deviations of previous calculations \cite{4a,4b,4c} on Ps-He
scattering from measurements at low energies \cite{2} could be a
consequence of the nonorthogonality defect and/or the inadequacy of the
correlation effect in exchange-dominated Ps-impact scattering, specially
at low energies.

To address this problem we choose to remove the nonorthogonality from the
exchange kernel of the momentum-space CCA equation by using a suitable
model potential.  The additional simplicity of the present exchange
potential makes it very attractive for performing dynamical calculation in
many-electron systems. The exchange model is shown to be readily
applicable to electron- and Ps-impact scattering problems.  In order to
test the generality and reliability of the exchange model we apply it to
electron-hydrogen ($e^-$-H) and electron-helium ($e^-$-He) scattering, in
addition to Ps-He scattering.

We present a theoretical study of ortho-Ps-He scattering employing a
three-Ps-state CCA scheme in momentum space where the usual nonorthogonal
exchange kernel arrising from antisymmetrization is replaced by the
present model exchange potentials.  The helium atom is always assumed to
be in its initial ground state and the Ps(1s), Ps(2s), and Ps(2p) states
are included in the coupled-channel calculation.  Being the lightest atom,
Ps is more vulnerable to excitation than the inert helium atom in Ps-He
scattering.  Also, the Ps-excitation thresholds are the lowest ones in
this system.  Hence, the present three-Ps-state model seems to be a
reasonable one to describe Ps-He scattering from low to medium energies.
The cross sections for higher discrete and continuum excitations of Ps
atom are calculated in the framework of the first Born approximation
including present exchange. These Born cross sections are added to the
above three-Ps-state cross sections to predict the target elastic total
cross section.

The plan of the paper is as follows. In Sec. II we present the model
exchange potential for electron-impact scattering and numerical results
for electron scattering by H and He.  In Sec. III we present the model
exchange potential for Ps-impact scattering and numerical results for Ps
scattering by He.  Finally, in Sec. IV we present a summary of our
findings.

\section{Exchange Potential for Electron-Impact Scattering}

Although, we are mostly interested in developing an exchange potential for
Ps-impact scattering in this work, first we illustrate and check our model
in the case of electron-atom scattering where the exchange potential is
well under control.  We develop the present exchange model for $e^-$-H
elastic scattering using H(1s) orbital and finally extend it to the case
of inelastic scattering by a complex target described by a Hartree-Fock
(HF) wave function.  The exchange potentials are derived from the
following exchange transition amplitude \begin{eqnarray} g({\bf k_f,k_i})
& = & -\frac{1}{2\pi} \int \makebox{d} {\bf r}_1 \makebox{d}{\bf r}_{2}
\phi^*({\bf r}_{2}) 
 \frac{1}{{r_{12}}}\phi({\bf r}_1)\nonumber \\ &\times& \exp[ \makebox{i}(
{\bf k_i.  r}_{2} - {\bf k_f. r}_1)] ,\label{1x} \end{eqnarray} where the
position vector of the incident (target) electron is ${\bf r}_{2}$ (${\bf
r}_1$).  Here $\phi$ is the wave function of H, ${\bf k_i}$ (${\bf k_f}$)
is the initial (final) momentum of the incident electron, and ${\bf
r}_{12}={\bf r}_1 -{\bf r}_{2}$.  Amplitude (\ref{1x}) is the leading term
of the exchange amplitude at large energies \cite{OR} and also the usual
starting point for deriving model exchange potentials \cite{T}.  To remove
the nonorthogonality defect we seek an exchange potential of the form
\begin{equation}\label{2x} g({\bf k_f,k_i}) \sim \int \makebox{d}{\bf r}
\phi^*({\bf r}) U({\bf r, k_i,k_f})  \phi({\bf r}), \end{equation} where
the form of $U$ is to be determined.  We consider the integration over the
coordinate of the final projectile electron ${\bf r}_1$ of Eq. (\ref{1x}) 
below.  Using $\phi ({\bf r})=\pi ^{-1/2}\alpha^{3/2} \exp(-\alpha r)$,
taking Fourier transformation, and performing the integration over ${\bf
r}_1$, we obtain \begin{eqnarray} I& \equiv & \int \makebox{d}{\bf r}_1
\frac{1}{r_{12}}\phi({\bf r}_1)\exp ( - \makebox{i}{\bf k_f. r}_1)
\nonumber \\ &=& \frac {{4\alpha}^{5/2}}{\pi^{3/2}}\int \makebox{d}{\bf q}
\frac{\exp(- \makebox{i}{\bf k_f.r}_{2})}{({\bf k_f}-{\bf q})^2}
\frac{\exp( \makebox{i}{\bf q.  r }_{2})}{(q^2+\alpha^2)^2}.\label{3x}
\end{eqnarray} Any average value prescription for $({\bf k_f}-{\bf q})^2$
in Eq. (\ref{3x})  will reduce Eq. (\ref{1x}) to form (\ref{2x}). Then, in
the model exchange potential, the final and initial state wave functions
will be expressed in terms of same coordinates.  Recalling that internal
kinetic energy of H $(q^2/2)$ is given by $\alpha^2/2$ in atomic units, we
take average of $q^2$ as $\alpha^2$, and set $({\bf k_f}-{\bf q})^2
\approx (k_f^2+\alpha^2)$, where the average value of the scalar product
is assumed to be zero.  After taking an inverse Fourier transformation in
Eq.(\ref{3x}), the final model exchange potential takes the following
simple
form \begin{eqnarray}\label{4x}
 g ({\bf k_f,k_i})&\approx &\frac{-2}{k_f^2+\alpha^2} \int \phi^*({\bf
r}_2)\exp ( \makebox{i} {\bf Q. r}_2)\phi ({\bf r}_2)  \makebox{d}{\bf
r}_2, \end{eqnarray} where ${\bf Q= k_i-k_f}.$ Although we derived Eq.
(\ref{4x}) for elastic scattering, this result is straightforwardly
extendable to inelastic $e^-$-H scattering to a final H(2s,2p),
H(3s,3p,3d), ..., etc. orbital. In such cases the final model exchange
potential for transition from state $\nu$ to $\nu {'}$ becomes
\begin{eqnarray}\label{5x}
 g_{\nu '\nu} ({\bf k_f,k_i})&\approx &- \frac{2}{k_f^2+\alpha_\nu  ^2}
\int \phi_{\nu '}^*({\bf r}_2)\exp ( \makebox{i}
 {\bf Q. r}_2)\phi_{\nu}({\bf r}_2)
\makebox{d}{\bf r}_2,\nonumber \\
\end{eqnarray}
where the parameter $\alpha_\nu$ refers to the initial state $\nu$.

Similar model potentials were derived by Ochkur and also by
Rudge \cite{OR}.  Ochkur's result is obtained by setting $\alpha = 0 $ in the
prefactor of Eq. (\ref{5x}). Rudge's result corresponds to taking the
prefactor ${(k_f^2+\alpha^2)}^{-1} =  (k_f-i\alpha)^{-2}$.  The model
exchange potential (\ref{5x}) has the following desirable physical
properties. This potential is the strongest at the lowest possible energy
($k_f=0$) for the  weakest bound atomic orbital ($\alpha_\nu  \to 0$).
Hence,
the effect of exchange is more pronounced at low energies for the loosely
bound orbitals.
 
For a general  HF wave function,  $\psi _ \nu ({\bf r}_1,...,{\bf
r}_j,\-...,{\bf r}_N)\- ={\cal A} [\prod _ {j=1} ^ N\phi_{\nu j} ({\bf r}_j)],
$ where ${\cal A}$  is the antisymmetrization operator and the position
vectors of the electrons  are ${\bf r} _j,$ $j=1,2,...,N$ and the atomic
orbitals $\phi_{\nu j}({\bf r})$ have the following form:
\begin{equation}
 \phi_{\nu j}({\bf r}) \equiv \sum_\kappa a_{\kappa  j} \phi_{
 \kappa 
j}({\bf r}),  \label{6x}\end{equation} 
where the index $\nu$ representing the atomic state is dropped on the
right-hand side.
Summing over appropriate target
electrons $j$ and allowing for inelastic channels, the full  exchange
potential is given by
\begin{eqnarray}\label{7x}
B_{E,\nu{'}\nu}({\bf k_f,k_i})&=&\sum_j g_j =-\sum_j \sum_{\kappa\kappa'}
\frac{2a_{ \kappa  j}
a_{\kappa' j}}{D_{\kappa\kappa '  j }}\nonumber
\\ &\times&
\int \phi_{\kappa'j}^*({\bf r}) \exp ( \makebox{i} {\bf Q. r})\phi_{ 
\kappa j}({\bf r}) 
\makebox{d}{\bf
r},
\end{eqnarray}
with 
\begin{equation}\label{a1}
D_{\kappa\kappa '   j}=[{k_f^2 + \alpha_{ \kappa j }^2}],
\end{equation}
where $\phi_{ \kappa j }({\bf r})$ is the   $\kappa$th function
 of the $j$th electron, and $\alpha_{ \kappa j}$ refer to the
initial state.

 The model potential (\ref{7x}) with prefactor $D_{\kappa\kappa '   j}$
  of                                          
Eq. (\ref{a1}) is not time-reversal symmetric. However, if we perform the      
  integration over the initial projectile electron ${\bf r}_2$ in Eq.       
(\ref{1x})                                                                  
first, and carry on a similar procedure, we obtain exchange potential      
(\ref{7x}) with  $D_{\kappa\kappa '   j}=(k_i^2+\alpha_{ \kappa'         
j}^2)$, where $\alpha_{\kappa ' j}$ refer to the final state.
 These two possibilities suggest the following                            
symmetric prefactor                                                           
\begin{equation}\label{a2}                                                    
D_{\kappa\kappa '   j}=[(k_i^2+k_f^2)/2+(\alpha_{\kappa                           
  j}^2+\alpha_{ \kappa' j}^2)/2]
    \end{equation}                 in Eq. (\ref{7x}).                                                  
  The two
possibilities (\ref{a1})  and (\ref{a2})  corresponding to two averaging
  At high energies, the model 
exchange potential (\ref{7x}) with different averaging
prescriptions lead to the Oppenheimer exchange potential \cite{Opp}. 
However, at low energies the cross section is sensitive to the 
averaging procedure and the value of the parameter $\alpha$ in
prefactors (\ref{a1}) or (\ref{a2}). This sensitivity may well be
exploited to tune the parameter $\alpha $ of a particular averaging 
procedure in
order to obtain a better fit with experiment at low energies. 

Although, the model potential (\ref{7x})  is derived for
the ground state of the atomic target, it is straightforward to see that
the same result is also valid for target excitations in the final state
using a similar averaging prescription. Hence model potential (\ref{7x})
is equally 
valid for both elastic and inelastic scattering by the target.

We have used the exchange potential (\ref{7x})  in  $e^-$-H and $e^-$-He
scattering and calculated the elastic cross sections. We also demonstrate
the effect of different averaging procedures $-$ symmetric and
nonsymmetric, and the variation of the parameter $\alpha$ whenever
relevant. In the case of $e^-$-H
scattering  we exhibit the results for elastic cross section in a coupled
H(1s,2s,2p)  model using the above exchange potential in  the symmetric
form (\ref{a2}) with the exact value of the parameter $\alpha$.  For
$e^-$-He scattering we present results for elastic cross section in the
static-exchange model  using the symmetric form (\ref{a2}). In the case of
He we present
a variation of the parameter so as to obtain a better fit with experiment.

In Fig. 1 we present results for $e^-$-H scattering using Eqs. (\ref{7x}) 
and (\ref{a2}), where we exhibit the exchange Born,  static-exchange, 
and H(1s,2s,2p) cross sections without variation of the parameter
$\alpha$. In this figure we compare the low-energy cross sections  with
experimental  results \cite{N} and the  calculations by Temkin and Lamkin
\cite{6a}. At medium energies the results are compared with essentially
converged
calculation of Callaway \cite{C}. We also plot  the total first Born 
cross section  with Oppenheimer exchange \cite{Opp}. At low energies the 
present H(1s,2s,2p) cross sections are  improvement over the present
static-exchange cross sections. At higher energies they are
essentially identical and only the static-exchange results are shown.  
 At large energies, as
expected, the present cross sections tend to the total first exchange-Born
(Born+Oppenheimer exchange) results.  Both at low and medium energies the
agreement of the present  cross sections with the results of
other workers is  encouraging. We verified that  both the
exchange Born,  static-exchange cross sections are sensitive 
to the 
variation of the parameter $\alpha$ in the prefactor (\ref{a2}).  We 
demonstrate the effect of such variation at low energies in the study of 
$e^-$-He scattering where it seems more relevant.

In Fig. 2 we plot the present static-exchange cross section of  
electron-helium  scattering for model exchange potential given by Eqs.
(\ref{7x}) and (\ref{a2}) with the HF helium wave function of Ref.
\cite{10a}.  In this case we present results for first
exchange-Born and static-exchange elastic cross sections with the exact
parameters  $\alpha$'s. Here we also present
results for  
static-exchange  cross sections 
with  modified value for the 
parameters $\alpha$ ($<q^ 2>= (0.4\alpha)^ 2$)
 in the prefactor (\ref{a2}) for both $\kappa$ and $\kappa ' $ corresponding
to initial and final states, respectively.
(The parameters in the helium wave function under the integral in Eq.
({\ref{7x}}) are left unchanged as should be.) The results are compared
with experimental results and the five-state
[He(1s,2$^1$s,2$^3$s,2$^1$p,2$^3$p)]  CCA
calculation of 
Burke et al \cite{bk} using model exchange potential.

The present static-exchange cross-sections 
 agree reasonably with experiment \cite{T1} at medium to high energies. 
The variation of the parameter $\alpha$ 
in this case has led to good agreement with experiment and the
CCA calculation of Burke et al \cite{bk} at lower energies.  
For obtaining a better agreement with experiment the effect of excitation
and polarization of the target should be taken into account. This could be 
done by considering a coupled-channel calculation with helium
excitations as in  the  electron-hydrogen  scattering considered above. 
With this reliability achieved in the $e^-$-H
and $e^-$-He systems we extend this exchange model to Ps impact cases.

\section{Exchange Potential for Positronium-Impact Scattering}

\subsection{Formulation}

Here, we first we develop the present exchange model potential for Ps-H
elastic scattering using H(1s) orbital and finally extend it to inelastic
Ps scattering by a many-body target described by a HF wave function.  We
start with the following  exchange transition amplitude \cite{T}
\begin{eqnarray} 
g ({\bf k_f,k_i})& = & 
-\frac{1}{\pi} \int \makebox{d}{\bf x} \makebox{d}
{\bf r}_1 \makebox{d}{\bf r}_{2}
\phi^*({\bf r}_{2})
\chi^*({\bf  t }_{1}) \frac{1}{{r_{12}}}\phi({\bf r}_1)
\nonumber \\ & \times & 
\chi({\bf  t }_{2})  \exp[ \makebox{i}( {\bf k_i.
s}_{2} - {\bf k_f. s}_1)] ,\label{1}
\end{eqnarray}
where  the position vector of the electron (positron) of Ps is ${\bf r}_{2}$
(${\bf x}$). Here ${\bf s}_j = ({\bf x}+{\bf r}_{j})/2,$ ${\bf  t }_{j}=({\bf
x}-{\bf r}_j)$,  $j=1,2$, $\chi$ ($\phi$) is the wave function of Ps (H).  As
in the previous section, to remove the nonorthogonality defect we seek an
exchange potential of the form
\begin{equation}\label{1a}
g({\bf k_f,k_i}) \sim  \int \makebox{d}{\bf r} \makebox{d} {\bf t}
\phi^*({\bf r})
\chi^*({\bf  t })U({\bf r,t, k_i,k_f})
\phi({\bf r})
\chi({\bf  t }),
\end{equation}
where  $U$ is to be determined.  We consider the  integration over the
coordinate of the final projectile electron ${\bf r}_1$ of Eq. (\ref{1})
below.  Using $\phi ({\bf r}) = \pi ^{-1/2}\alpha^{3/2} \exp(-\alpha r)$ and
$\chi ({\bf t})=\pi ^{-1/2}\beta^{3/2} \exp(-\beta t)$, taking Fourier
transformation, the integral $\cal I$ over ${\bf r}_1$ is given by
\begin{eqnarray}
\cal I& \equiv &  \int  \makebox{d}{\bf r}_1 
\chi^*({\bf  t }_{1}) \frac{1}{{r_{12}}}\phi({\bf r}_1)
  \exp(- \makebox{i}  {\bf k_f. s}_1) ,\nonumber \\ &=&
\frac {{4(\alpha\beta)}^{5/2}}{\pi^4}\int \makebox{d}{\bf p}\makebox{d}{\bf q}
\frac{\exp(- \makebox{i}
{\bf k_f.r}_{2}/2)}{({\bf k_f}/2-{\bf p+q})^2}\nonumber \\ &\times&
\frac{\exp( \makebox{i}{\bf q.  t }_{2})}{(q^2+\beta^2)^2}
\frac{\exp( \makebox{i}{\bf p.r}_{2})}{(p^2+\alpha^2)^2}.\label{5}
\end{eqnarray} 
Again we employ an average value prescription for  $({\bf k_f}/2-{\bf
p+q})^2$ in Eq. (\ref{5})  which will reduce Eq. (\ref{1}) to form
(\ref{1a}).  Recalling that the
internal kinetic energies of H (represented by $p^2/2m;$ $m=1$)  and Ps
($q^2/2m$; $m=1/2$) are given by $\alpha ^2/2 $  and $\beta ^2$ in atomic
units, we take the averages  of $p^2$ and $q^2$  as $\alpha^2$ and $\beta^2$,
respectively, and set  $({\bf k_f}/2-{\bf p+q})^2  \approx
(k_f^2/4+\alpha^2+\beta^2)$ in Eq. (\ref{5}), where the average values of the
scalar products are assumed  to be zero.
  After taking an inverse Fourier transformation in
Eq.(\ref{5}) and transforming the set of variables ${\bf x, r}_1, {\bf
r}_2$ to ${\bf t}_2, {\bf r}_1, {\bf r}_2$, where the Jacobian is unity, 
the final model exchange potential becomes
\begin{eqnarray}\label{6}
 g ({\bf k_f,k_i})&\approx &- \frac{4(-1)^{l'+1}}{k_f^2/4+\alpha^2+\beta^2}
\int \phi^*({\bf r}_2)\exp ( \makebox{i} {\bf Q. r}_2)\phi({\bf r}_2)
\makebox{d}{\bf r}_2\nonumber \\ &\times&
\int \chi^*({\bf  t }_2)\exp ( \makebox{i}{\bf Q}.{\bf  t }_2/
2)\chi({\bf  t }_2 ) \makebox{d}{\bf  t }_2,
\end{eqnarray}
where $l'$ is the angular momentum of the final Ps state and 
 Eq. (\ref{6}) has been multiplied by $(-1)^{l'+1}$
the final-state parity. 
This provides the correct sign of the
exchange potential given by formal antisymmetrization for elastic and all Ps
excitation channels.  This exchange potential could be considered to be a
generalization of Rudge-type exchange Born amplitude \cite{OR} for
electron-impact scattering to more complex situations.  For a general  HF
orbital (\ref{6x}), summing   over appropriate target electrons $j$ and
allowing for inelastic Ps channels, the (target-elastic)  model exchange
potential is given by
\begin{eqnarray}\label{7}
B_{E,\mu'\mu}({\bf k_f,k_i})&=&\sum_j g_j=-\biggr[
\sum_j \sum_{\kappa \kappa'}
\frac{4a_{\kappa j} a_{\kappa' j}(-1)^{l'+1}}{D_{\kappa\kappa '   j}} \nonumber
\\ &\times&
\int \phi_{\kappa'j}^*({\bf r}) \exp ( \makebox{i} {\bf Q. r})\phi_{\kappa j}
({\bf r}) \makebox{d}{\bf r}\biggr]
\nonumber \\ &\times& \int \chi^*_{n' l'}({\bf  t })\exp ( \makebox{i}{\bf
Q}.{\bf t }/2)\chi_ {n l}({\bf  t }) \makebox{d}{\bf  t },
\end{eqnarray}
with 
\begin{equation}
D_{\kappa\kappa '   j}= [{k_f^2/4+\alpha_{\kappa j}^2+
\beta_{n '} ^2}]\label{a}
\end{equation}
where $\mu\equiv nl$ ($\mu'\equiv  n'l'$) are the initial (final) Ps quantum
numbers, $\phi_{\kappa j} ({\bf r}) $ is the $\kappa$th function of the $j$th
electron for the atomic ground state, and $\beta _{n'}$ corresponds to the
final inelastic Ps state, for which the derivation of the model potential is
similar and   leads to the same result (\ref{6}) or (\ref{7}).  For Ps
ionization, the constant $\beta_{n'}^2$, which corresponds to the final
Ps-state binding energy, is taken as 0 in Eq.  (\ref{7}).

As noted in Sec. II,
the exchange potential given by Eqs. (\ref{7}) and (\ref{a}) is not
time-reversal symmetric. However, if one performs in  Eq. (\ref{1}) the
integration over the coordinate of the initial projectile electron 
${\bf r}_ 2$ first with a similar average-value prescription as above
one will obtain exchange potential (\ref{7}) with 
$D_{\kappa\kappa '   j}= [{k_i^2/4+\alpha_{\kappa ' j}^2+
\beta_{n } ^2}]$. These two possibilities suggest the following symmetric 
prefactor
\begin{equation}
D_{\kappa\kappa '   j}= [(k_f^2+k_i^2)/8+(\alpha_{\kappa
j}^2+\alpha_{\kappa ' j}^2)/2+
(\beta_{n '} ^2+\beta_{n } ^2)/2].\label{b}
\end{equation}
Both choices (\ref{a})  and (\ref{b}) lead to good numerical results. 
At high energies the results are independent of this choice. At low
energies they are sensitive to the choice and the value of the parameter 
$\alpha $ in Eqs. (\ref{a})  and (\ref{b}). 
In this work we shall present only results of choice (\ref{a}) 
with the original and modified value of the parameter $\alpha$.

The target-elastic direct Born Ps-He amplitude for Ps transition from 
state $\mu$ to $\mu '$ is given by \cite{4d} \begin{eqnarray}\label{8}
B_{D , \mu '\mu}&(&{\bf k_f,k_i})= \frac {4}{Q^{2}}  \biggr[2   -
\sum_{\kappa \kappa'}\sum_j a_{\kappa j}  a_{\kappa 'j}  \nonumber \\
&\times& \int \phi_{\kappa' j}^*({\bf r}) \exp (i {\bf Q. r})\phi_{\kappa
j} ({\bf r}) \makebox{d}{\bf r} \biggr] \nonumber \\ &\times &  \int
\chi^*_{\mu '}({\bf  t }) [\exp (\makebox{i}{\bf Q}.{\bf t }/2) - \exp
(-\makebox{i}{\bf Q}.{\bf t  }/2)]\chi_ {\mu}({\bf  t }) \makebox{d}{\bf
t }\nonumber \\ \end{eqnarray}
With the present prescription, the Ps-impact exchange potential is
written in the
form of
product of projectile and target form factors, as the   
  direct  potential above. 
This simple form of the amplitudes
facilitates numerical calculations.

\subsection{Numerical Application to Ps-He Scattering}

In the case of target-elastic Ps-He scattering, electron exchange between
the incident Ps and target He is only possible between like spins.
Consequently, only the spin-triplet state of the electrons undergoing
exchange is possible. We  define appropriately symmetrized spin-triplet
``Born" amplitudes, ${\cal
B}$,   via $ {\cal B}_{\mu '\mu}({\bf k_f,k_i}) = B_{D,\mu '\mu}({\bf
k_f,k_i})- B_{E,\mu '\mu}({\bf k _f,k_i}).$ The appropriately symmetrized
scattering amplitude $f$ satisfies the following  momentum-space 
Lippmann-Schwinger scattering integral 
equation \cite{4c}
\begin{eqnarray}
f_{\mu '\mu}({\bf k',k})&=&{\cal B}_{\mu '\mu}({\bf k ',k}) \nonumber \\
&-&\sum_{\mu ''}\int \frac{\makebox{d}{\bf k ''}}{2\pi^2}\frac {{\cal B}_
 {\mu '\mu ''}({\bf k ',k''}) f_{\mu''\mu}({\bf k'',k}) }
{E-\epsilon_{\mu ''}-k ''^2/4+ \makebox{i}0},\nonumber \\ \label{z}
\end{eqnarray} where $\epsilon_{\mu''} $ is the total energy of the Ps
and He states in the intermediate state $\mu ''$ and $E$ is the total 
energy of the system.  
The differential cross section is defined by
$(d\sigma/d\Omega)_{\mu ', \mu} = (k'/k )|f_{\mu '\mu}({\bf k',k})|^2$.

We  performed static-exchange [with $\mu ''$ = Ps(1s) in Eq. (\ref{z})] and
three-Ps-state [with $\mu ''$ = Ps(1s,2s,2p) in Eq.  (\ref{z})] calculations
using exact wave functions for Ps and  HF atomic orbitals for He
\cite{10a}.  After a partial-wave projection, Eq. (\ref{z}) was
solved by the method of matrix inversion. Maximum number of partial waves
included in the calculation was 10.  Contribution of higher partial waves to
cross sections was included by corresponding Born terms. 
To predict the
cross sections at medium energies, we also calculated the discrete excitation
(3s, 3p, 3d, 4s, 4p, 4d, 4f, 5p, 5d, 5f, 6p) and ionization cross sections
of Ps in
the first Born approximation  keeping the target
frozen to its initial ground state using the present exchange model.

In Fig. 3 we plot the present target-elastic total cross section
[Ps(1s,2s,2p) three-Ps-state cross section plus target-elastic total Born
cross sections for $n \ge 3$ Ps-excitations and Ps-ionization]. 
  The experimental total Ps-He cross sections of two different groups 
  $-$ recent low-energy  cross section of Ref. \cite{2}
and medium- to high-energy cross sections of Ref. \cite{1} $-$
 are  also plotted.   For comparison we also plot the static-exchange and
22-coupled-pseudo-state  (without exchange) cross sections of Refs.
\cite{4c}  and \cite{4d}, respectively.  The measured Ps-impact total
cross section of Ref. \cite{1} has a peak near 20 eV and a lowering trend
below this energy, and the recent measurement around 1 eV of  Ref.
\cite{2} is consistent  with this trend. However, due to large error bar 
of the measurement of Ref. \cite{1} at their lowest energy (10 eV)
and due to inadequate data in this energy region, it is 
not clear from experiment whether the total cross section has a minimum
near the Ps excitation threshold or not. This question is addressed in the 
present theoretical investigation. At energies below the Ps-excitation 
thresholds, the elastic cross section is found to be a monotonically 
decreasing function of energy, as is usually found in many similar
scattering problems. In the narrow energy band between 5.1 to 6.8 eV, 
all the  Ps-inelastic channels open up causing a sharp increase of the 
total cross section, as can be seen in Fig. 7,  resulting in a minimum of
total cross section near the Ps-excitation threshold. With this feature of
the cross section, the present calculation 
bridges the two different
experimental findings and points out a minimum in total cross section near
the Ps(2s) threshold.  This feature is also noticed in the unpublished
theoretical work of Peach \cite{gp}. 
While the 22-coupled-pseudo-state calculation
\cite{4d}, which includes the Ps  excitation and ionization effects
through pseudo states,  completely denies this trend; the static-exchange
cross sections \cite{4a,4b,4c} are too large to match the measurement near
Ps(2s) threshold. 

So far we have parametrized the model potential from a physical argument
and presented results with it.   In Eq. (\ref{7}) $\alpha_{\kappa j}$'s
are parameters of  HF orbitals. In Fig. 3 we also exhibit the consequence
of a small variation of $\alpha_{\kappa j}$ in the prefactor
$(k_f^2/4+\alpha_{\kappa j}^2+\beta_{n'}^2)^{-1}$ of Eq. (\ref{7}).  The full
line, providing an overall better agreement with experiment,  is obtained by
varying parameters $\alpha_{\kappa j}$'s in the prefactor (\ref{a}), which is
taken as
\begin{equation}\label{c}
D_{\kappa\kappa '   j}= [{k_f^2/4+(0.88\alpha_{\kappa j})^2+
\beta_{n '} ^2}]\end{equation}
in both the static-exchange and three-Ps-state calculations.  Unless
specifically mentioned, all results presented here are calculated with this
modified prefactor.  The above reduction in the average  value of $<p^ 2>$
has led to a better agreement with experiment.

Next we present an account of  phase shifts and angle-integrated partial
cross sections with modified prefactor (\ref{c}).  The present
static-exchange and three-Ps-state elastic scattering phase shifts for
different partial waves  below the lowest excitation threshold are shown in
Figs. 4 (S wave) and 5 (P and D waves).  The present phase shifts are 
different from those of previous calculations
\cite{4a,4b,4c} as is expected from the cross-section pattern. However, for
comparison we show the phase shifts of the recent work by Sarkar and Ghosh
\cite{4c} in Figs. 4 and 5.  At these energies the S-wave phase shifts alone
control the elastic cross section. The present low-energy elastic S-wave
phase shifts are expected to be more reasonable to those of the previous
calculations as from Fig. 3 we find that the present cross sections are in
better agreement with experiment.

In Fig. 6 we plot the present low-energy elastic cross sections for  
static-exchange
and three-Ps-state calculations. We compare these cross sections with the
recent sophisticated 
low-energy experimental cross section
  of Ref. \cite{2} measured using time-resolved Doppler Broadening 
 Spectroscopy  and the previous  static-exchange
cross sections of Refs. \cite{4a,4b,4c}. 
Here, just to have a
feeling, we also plot the present static-exchange cross section calculated
with the exact parameter $\alpha$ in the prefactor. The present 
three-Ps-state cross sections 
are significantly smaller than previous theoretical cross sections and 
are in close agreement with experiment. 
The present exchange Born cross sections are also smaller than those of
previous calculations.  
For example, Sarkar and Ghosh \cite{4c} obtained the
first Born  elastic cross sections 131.9$\pi a_0^2$  and 12.5$\pi a_0^2$ at
0.068 eV and 1 eV,
 compared to the present first Born  elastic cross sections
8.8$\pi a_0^2$ and  3.7$\pi a_0^2$, respectively. 
The present three-Ps-state cross sections
are  smaller than the static-exchange cross sections
and  are well within the experimental error bar.  
However, in the present 
calculation we have neglected long-range van der Waals interaction.
The effect of this interaction has been shown \cite{4b} to increase the 
zero-energy cross section by as much as 30$\%$, while around 1 eV its effect
is to increase the cross section by about 3$\%$ only. Hence, the inclusion
of this  interaction is expected to further improve the agreement with
experiment at low energies.

In Fig. 7 we exhibit the different  angle-integrated partial cross sections
for the three-Ps-state  calculation. Here we show the Ps(1s) and Ps(2s+2p)
cross sections from the three-Ps-state calculation and Ps($n\ge 3$)
excitation and Ps-ionization cross sections using the present total first
Born model.  At medium energies the Ps-ionization cross sections are the
largest and dominates the trend of the total cross section of Fig. 3.  This
feature has also been observed by Campbell et al \cite{4d} in Ps-H
scattering.

\vskip .2cm 
{Table I: Angle-integrated Ps-He partial cross sections in $\pi a_0^2$ at
different positronium energies:  EB $-$ first Born with present
exchange; PSE $-$ present static exchange; SE $-$ static exchange of Ref.
\cite{4c};
TPS $-$ three-Ps-state with
present exchange}

\vskip .2cm

\begin{centering}

\begin{tabular} {|c|c|c|c|c|c|c|c|c|c|c|} \hline Energy  & Ps(1s) & Ps(2s)
&
Ps(2p) & Ps(1s)&Ps(1s) &
 Ps(1s) & Ps(2s)  & Ps(2p) & Ps($n\ge 3$) & Ps-ion \\
(eV) & EB & EB & EB & SE&PSE  & TPS & TPS & TPS &
EB  & EB\\
  \hline
 0.068 & 13.73 & & &14.4 &3.73 & 2.70&  & & & \\
 0.612 & 10.88& & & 12.9&3.34 &2.36 &  & &  &  \\
 1.088 & 9.05& & & 12.1&3.07&2.13 &  &  & &  \\
1.7 & 7.31& & &11.3&2.80 &1.88 & & & &  \\ 
2.448 &5.79& & & 10.5&2.52&  1.62 & & & &  \\ 
4.352 & 3.57& & & 9.0&1.99&1.09  & & & & \\
5 & 3.10& & & &1.85&0.89     & & & & \\
5.508 &2.81 &0.80($-1$) &1.51 & &1.75 &0.81 &0.49($-1$) &0.83 &  & \\
6 & 2.56& 0.10&1.87 & &1.66 &0.81         &0.70($-$1) &1.16 & & \\
6.8 & 2.22& 0.12& 1.98&7.7&1.53 &0.81 &0.74($-$1) &1.39 &0.69 &  \\
8 &1.84&0.11&1.86& & 1.35 &0.79 &0.64($-$1) & 1.44&  0.86    &0.74  \\
10 & 1.39 & 0.91($-1$) & 1.54 & & 1.11 &0.73 &0.52($-$1) & 1.31 &0.78  &2.05\\
15 &0.80&0.54($-1$)&1.00& & 0.71 &0.55 & 0.45($-1$) &  0.92& 0.52 & 3.67 \\
20 & 0.52&0.35($-1$)&0.72&3.6&0.49 &0.41 &0.33($-1$) &0.68 & 0.38 & 4.10 \\
30 &0.27&0.17($-1$)&0.44&2.0&0.26 &0.24 &0.18($-1$) &0.43 & 0.23  &  3.96\\
40 &0.16&0.10($-1$)&0.31 &0.7&0.16 &0.15 &0.11($-1$) &0.30 & 0.16 &  3.52  \\
50 & 0.11&0.65($-2$)&0.23& &0.11 &0.10 &0.68($-2$) &0.23 & 0.11 & 3.09 \\
60 &0.75($-1$)&0.45($-2$)&0.19&0.8($-1$)&0.75($-1$)&0.72($-1$) &0.46($-2$)&
0.18 &0.93($-$1)&2.72 \\
80 &0.41($-1$)& 0.24($-2$)&0.13&0.1($-1$)&0.41($-1$) & 0.40($-1$)&0.25($-2$)  &
0.13
&0.63($-$1) & 2.14 \\
100 &0.25($-1$)&0.14($-2$)&0.98($-1$)&0.2($-2$)&
0.25($-1$) &0.25($-1$) &0.14($-2$)&0.98($-1$)
&
0.49($-$1) & 1.74 \\
\hline
\end{tabular}

\end{centering}
\vskip 0.2cm

The angle-integrated partial cross sections are of crucial importance 
 and are presented in Table I. 
These partial cross sections  are 
calculated with the modified
prefactor (\ref{c}) and leads to total cross section 
in better agreement with experiment. These cross
sections  should be considered to be the most realistic  
results of the present model study except near zero energy where van der
Waals force might play a crucial role, which is not taken into account. 
In addition to the three-Ps-state cross sections we also present our
first Born and static-exchange results in Table I with modified prefactor
(\ref{c}). 
For comparison we also show the static-exchange results of Sarkar and 
Ghosh \cite{4c}.
The present elastic Born cross sections are much smaller than those of Ref.
\cite{4c}.  The present three-Ps-state elastic  cross sections 
are smaller  than the static-exchange  cross sections, which demonstrates the
effect of large polarizability of Ps.

\section{Summary}

We have presented simple model exchange potentials for electron- and
Ps-impact scattering suitable for performing dynamical calculation in
many-electron systems and checked it in  electron scattering from H and He
and applied it to Ps scattering from  He.  The present static-exchange and 
three-Ps-state coupled-channel cross
sections of
electron-impact scattering are in agreement with other existing results
\cite{6a,N,C,T1,bk}.  We
have performed static-exchange and three-Ps-state  calculations for  Ps-He
scattering at low and medium energies.  To exhibit the usefulness of the
present exchange at medium energies, higher excitations and ionization of Ps
are calculated using the first Born model with present exchange.  The
present target-elastic total cross sections agree well with experiment
\cite{1,2} both at  low and   medium energies.

The work is supported in part by the Conselho Nacional de Desenvolvimento
Cient\'\i fico e Tecnol\'ogico and Funda\c c\~ao de Amparo \`a Pesquisa do
Estado de S\~ao Paulo of Brazil.

\newpage

{\bf Figure Caption:}

1. Elastic electron-hydrogen cross section: 
present static exchange model (dashed-double-dotted line);
present Born (dashed line);
present H(1s,2s,2p) model
(full line); total first Born with Oppenheimer exchange
(dashed-triple-dotted line); polarized orbital model by Temkin and Lamkin
\cite{6a} at lower energies ($< 10$ eV) and CCA model by Callaway \cite{C} 
at higher energies (dashed-dotted line); experiment (solid circles, Ref.
\cite{N}) [H(1s,2s,2p) CCA model results by Burke and Schey  \cite{6a} are
very close to Temkin and Lamkin and are not shown].

2. Elastic electron-helium cross section:
present static exchange with exact parameters $\alpha$ (dashed-dotted
line); present Born with exact $\alpha$ (dashed line);
present static exchange with modified $\alpha$
(full line); He(1s,2$^1$s,2$^3$s,2$^1$p,2$^3$p) CCA calculation of Burke
et al.
(dashed-double-dotted
line, Ref. \cite{bk})
experiment (solid circles and crosses, Ref.
\cite{T1}).

3. Total Ps-He cross sections  at different positronium
energies: present target-elastic result  
 from three-Ps-state model
plus present  first exchange  Born for $n\ge 3$ excitations and
ionization of Ps (dashed 
line); present
target-elastic result with modified parameter $\alpha_{\kappa j}^2$
in the prefactor  
 (full line); static-exchange  model of
Sarkar and Ghosh (dashed-dotted line, Ref. \cite{4c}); 
22-coupled-pseudo-state model of McAlinden et al. (dashed-double-dotted
line, Ref. \cite{4d});
experiment
(square, Ref.
\cite{2}; circle Ref. \cite{1}).

4. S-wave elastic Ps-He phase shifts 
at different positronium energies: present three-Ps-state model (full line);
present static-exchange model (dashed line); static-exchange model 
of Sarkar and Ghosh (dotted line, Ref. \cite{4c}).

5. P- and D-wave elastic Ps-He phase shifts 
 at different positronium energies: notations are the same as in Fig. 4.

6. Angle-integrated Ps-He elastic cross section at low positronium
energies: present three-Ps-state  model (full line); present static-exchange 
model (dotted line);  present static-exchange with unmodified parameter 
$\alpha$ (dashed-tripple-dotted line);
Fraser (dashed-double-dotted 
line, Ref. \cite{4a}); Barker and Bransden
(dashed-dotted line, Ref. \cite{4b}); 
Sarkar and Ghosh (plus, Ref.
\cite{4c}); experiment shown by square (Ref. \cite{2}).

7. Angle-integrated Ps-He partial cross sections at different 
 positronium energies with exact $\alpha$:
present elastic  from three-Ps-state model
(full line) and Ps(2s+2p) excitation (dashed-dotted line)
from three-Ps-state model, present Ps($n \ge 3$) excitation
(dashed-double-dotted
line) and  Ps ionization (dashed line) using first Born approximation
with present exchange.

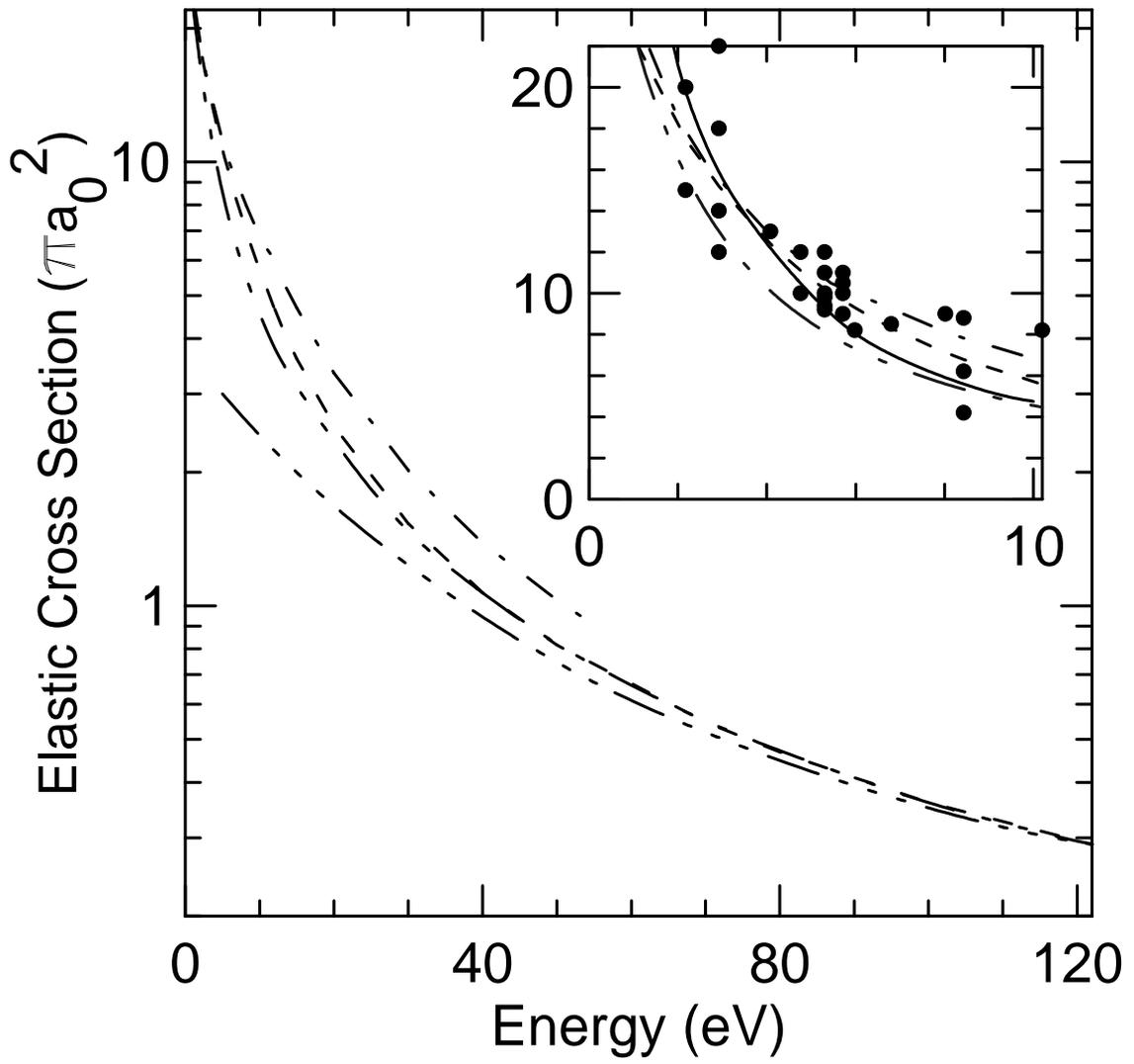
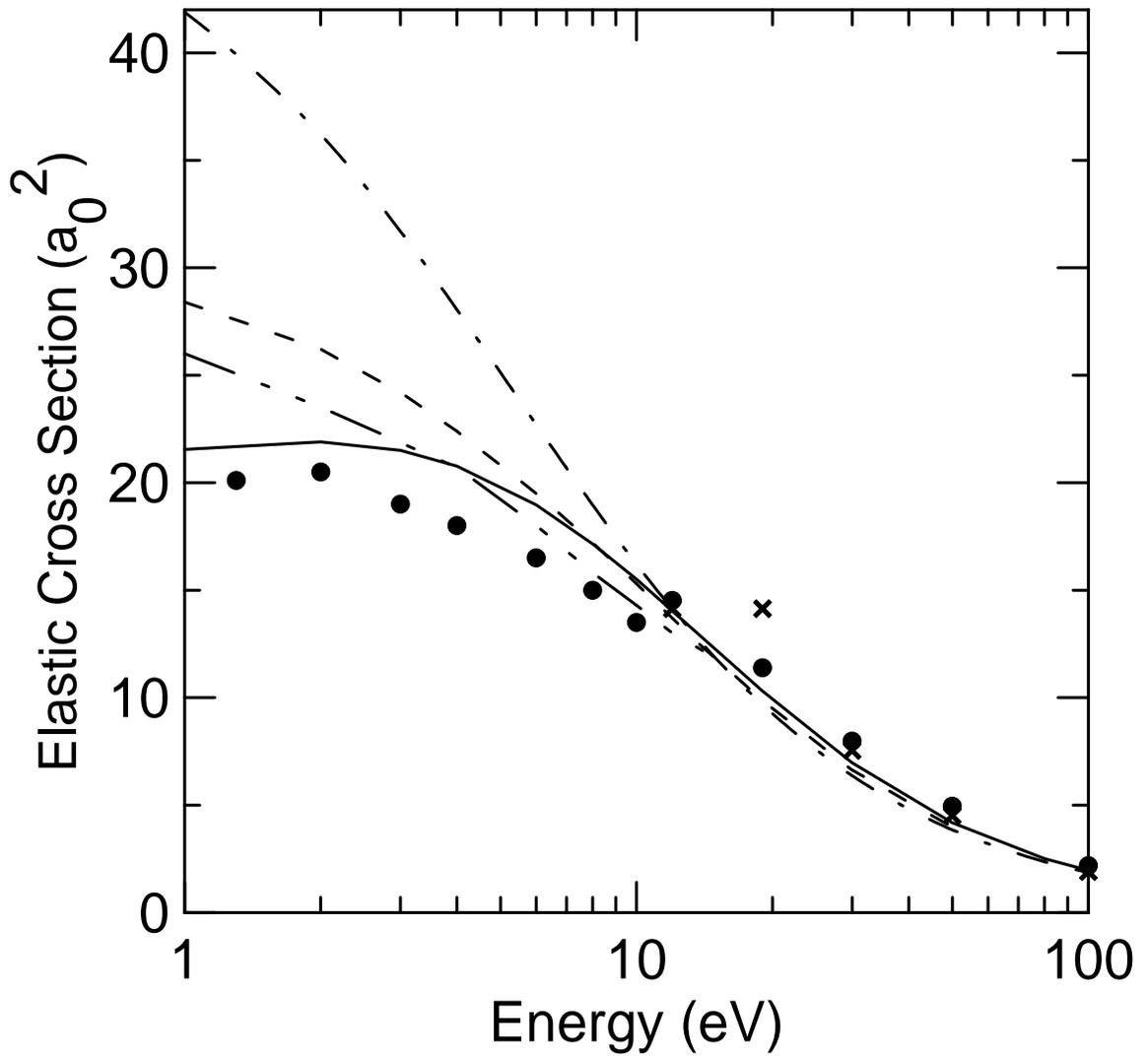
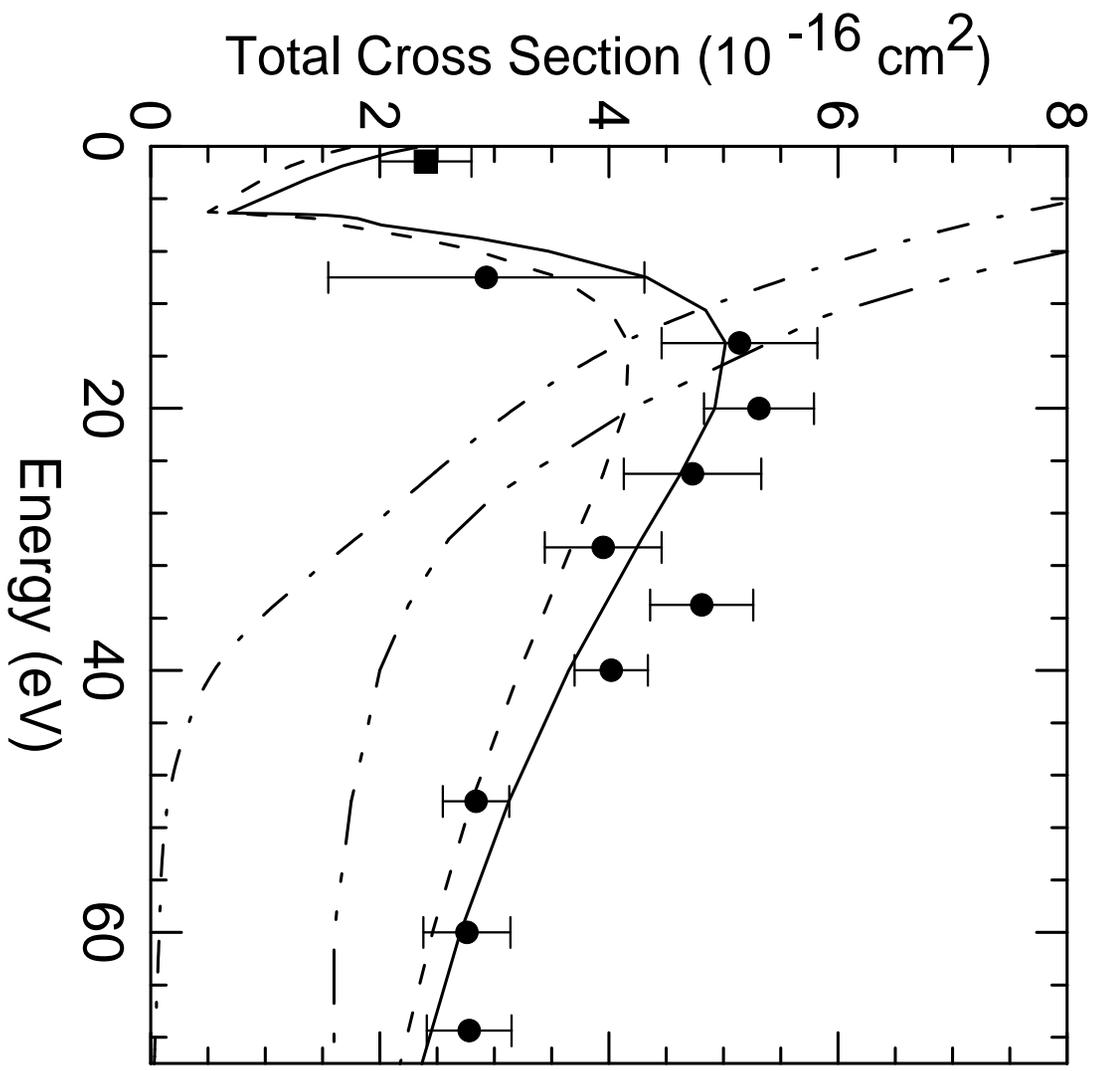
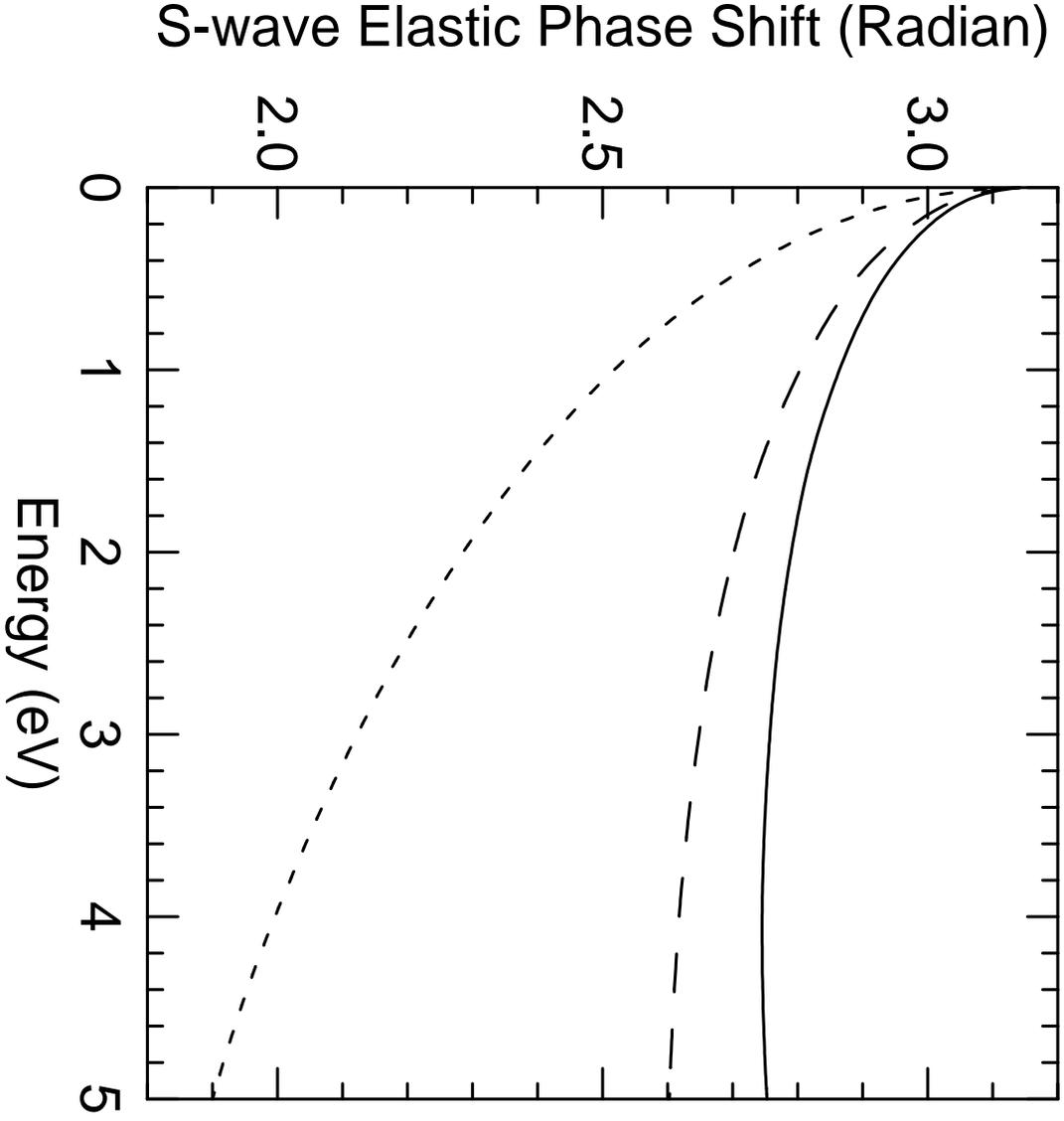
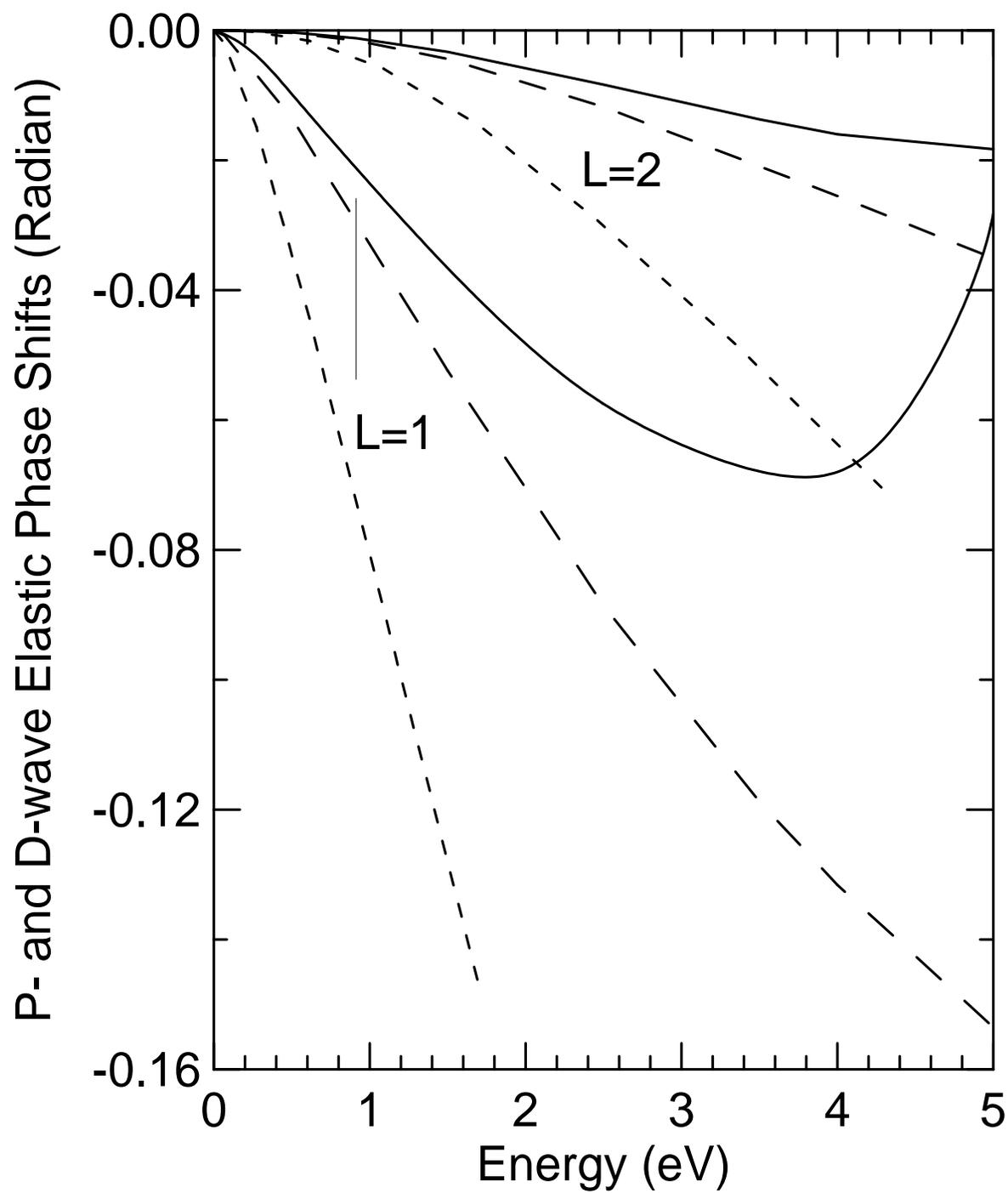
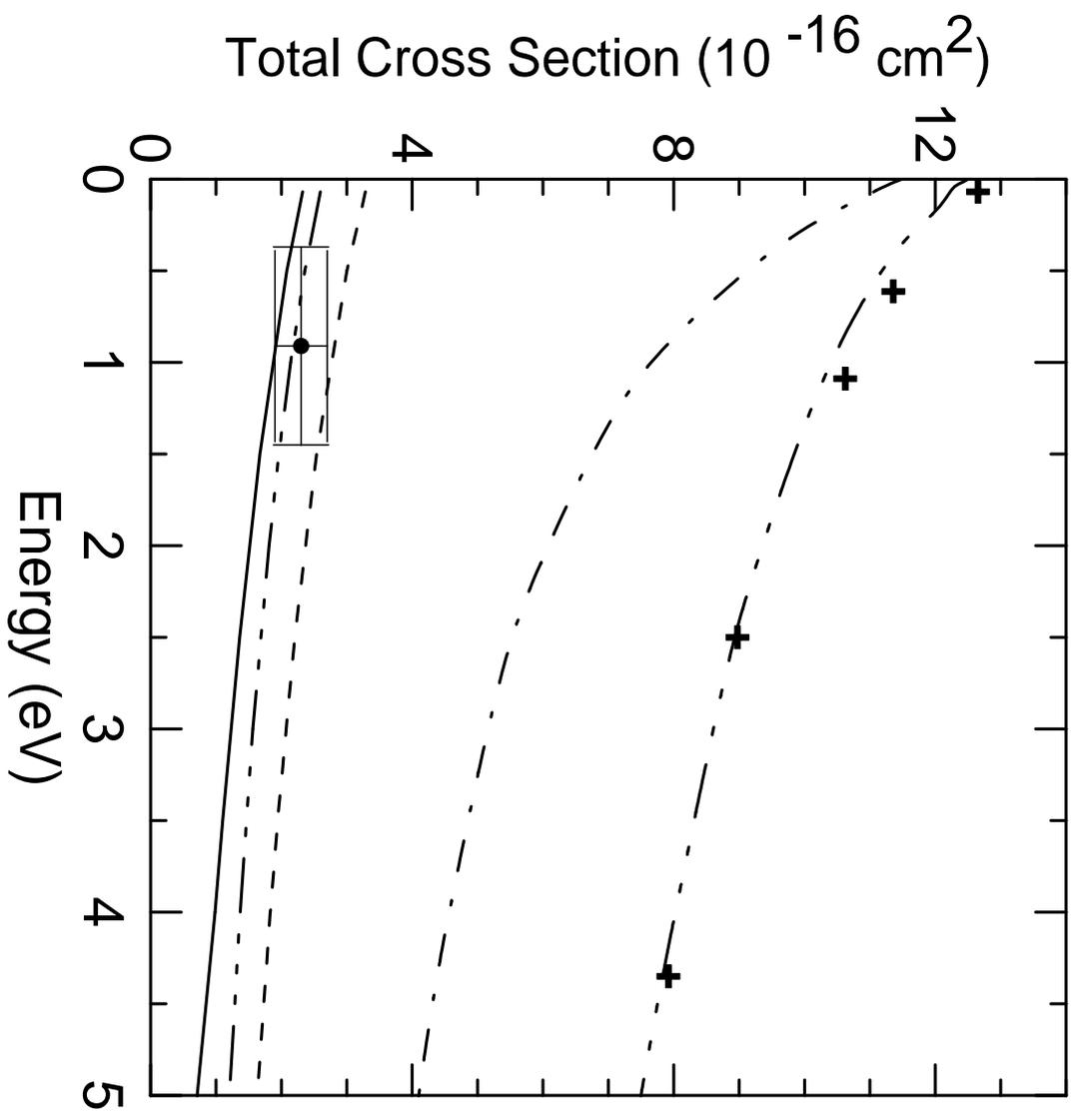
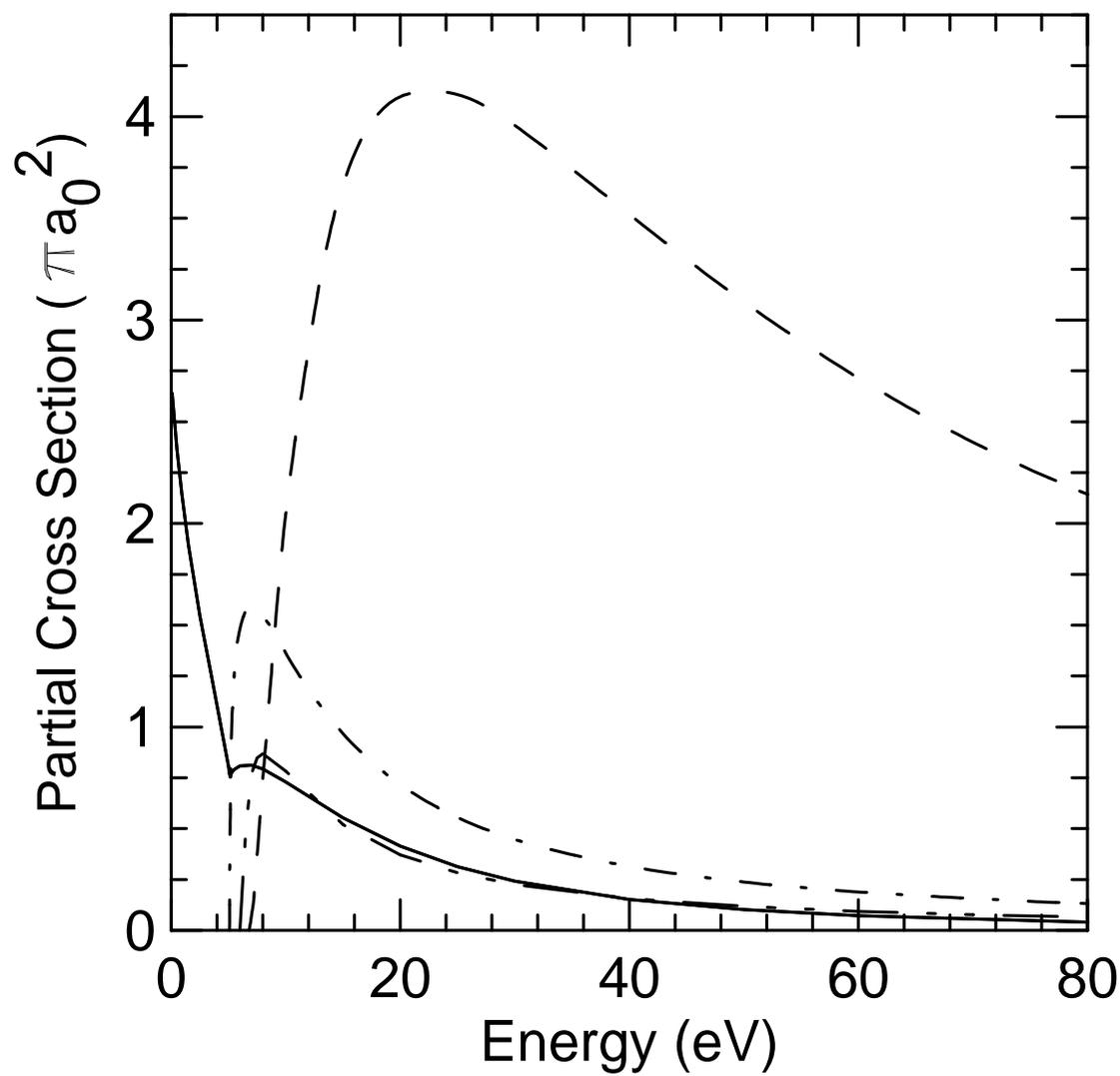
\end{document}